# Challenging targets or describing mismatches?
## A comment on Common Decoy Distribution by Madej et al.

Lucas Etourneau and Thomas Burger*
Univ. Grenoble Alpes, CNRS, CEA, Inserm, ProFI FR2048, Grenoble, France
* thomas.burger@cea.fr

**Abstract**: In their recent article, Madej et al.[1] proposed an original way to solve the recurrent issue of controlling for the false discovery rate (FDR) in peptide-spectrum-match (PSM) validation. Briefly, they proposed to derive a single precise distribution of decoy matches termed the Common Decoy Distribution (CDD) and to use it to control for FDR during a target-only search. Conceptually, this approach is appealing as it takes the best of two worlds, *i.e.*, decoy-based approaches (which leverage a large-scale collection of empirical mismatches) and decoy-free approaches (which are not subject to the randomness of decoy generation while sparing an additional database search). Interestingly, CDD also corresponds to a middle-of-the-road approach in statistics with respect to the two main families of FDR control procedures: Although historically based on estimating the false-positive distribution, FDR control has recently been demonstrated to be possible thanks to competition between the original variables (in proteomics, target sequences) and their fictional counterparts (in proteomics, decoys). Discriminating between these two theoretical trends is of prime importance for computational proteomics. In addition to highlighting why proteomics was a source of inspiration for theoretical biostatistics, it provides practical insights into the improvements that can be made to FDR control methods used in proteomics, including CDD.

## 1. A short history of FDR in biostatistics and proteomics

A False Discovery Rate (FDR) is a statistical estimate of the expected proportion of features that pass a significance threshold by chance (a.k.a., false discoveries). With the advent of high-throughput analyses, the number of measurable features have sky-rocketed. To avoid producing a proportional increase in false discoveries, it has become essential to control for the FDR (i.e., to conservatively select features based on the FDR). Although the starting point of FDR theory is unquestionably dated to 1995 with the publication of the seminal article by Benjamini and Hochberg (BH)[2], few later publications acknowledge the importance of pre-existing work.[3,4] After few technical improvements[5,6] between 1995 and 2000, the subject really gained momentum with the publication of the human genome[7], which revealed how high-throughput biology could dramatically take advantage of these hitherto purely theoretical advances. The early 2000s thus saw the emergence of several innovations. On the theoretical side, a group of researchers from Stanford reformulated the BH framework to better fit applications in biostatistics[8–12]. This notably led to the now well-established concepts of q-value (or adjusted p-value) and empirical null estimation.

Meanwhile, in the proteomics community, questions akin to FDR estimation showed up under several names (e.g., "false identification error rates"[13] in 2002 and "false-positive identification rate"[14] in 2003). It also coincided with the moment when Elias and Gigy[15] formulated their intuitions about false positive simulation through decoy permutations, preceding what is now known as Target-Decoy Competition[16] (TDC). It should be noted that this was a complete conceptual breakthrough at the time, as there was no statistical theory to support the idea that fictional variables (i.e., decoy sequences) created from the variables ones (i.e., target sequences) could be used to control for FDR. This is also why decoy databases were soon proposed for use in ways that were more compliant with the pre-existing theory of FDR control. Notably, in 2007-2008, two groups independently proposed that target and decoy searches be performed on separate databases, i.e., without organizing a competition between them (hereafter referred to as TDwoC, to emphasize the absence of competition).[17,18] The first, shorter and more conceptual, article[17] became the benchmark (despite the fact that the estimator proposed was far from optimal).[19] This article notably established the theoretical exactness of TDwoC by linking the approach to empirical null estimation,[9,12] a concept to which one of the co-authors had also contributed. In addition, they raised concerns about TDC in the conclusions of the article, as in their opinion, the additional competition procedure made it difficult to derive the distribution of target mismatch scores (a.k.a. target null PSMs). However, despite this warning as well as rare voices pointing out the apparent inaccuracy of TDC[20], the TDC approach progressively became the reference method over the course of the following decade.

This gap between practical approaches to FDR in proteomics and theoretical background in biostatistics was tentatively filled by He et al. (in works that remained largely unpublished[21,22] until recently[23]). Briefly, these authors demonstrated that FDR could be controlled (at the peptide-only level, as opposed to the more classically-considered PSM level) using decoy sequences. They connected their demonstration to simultaneously emerging studies from the Candès group[24]. Although Barber and Candès' seminal work unleashed an important and on-going renewal of FDR theory in the statistics community[25–28], it may seem old to proteomics researchers, as its core idea is to fabricate uninteresting putative biomarkers *in silico* (i.e., fictional variables referred to as "knockoffs") and to use them to challenge each real putative biomarker through pairwise competition.

## 2. Two distinct approaches to FDR

Today, the FDR can be controlled in two ways, both in theoretical statistics and in proteomics: based on a description of how false-positives distribute; or based on competition, by challenging the variables of interest with fictional ones (a.k.a. knockoffs or decoys). We hereafter summarize the two trends, along with their specificities.

### 2.1. "Describing decoys" or the null-based approach

The oldest approach is based on a simple rationale: The scores of observations we are not interested in (spectrum/peptide mismatches) form the so-called null distribution in statistics. If enough is known about the null distribution, then it is possible to "subtract" it from the distribution observed. We will be left with observations that lie beyond the null distribution, which can therefore be considered of significant interest (discoveries); in sum, to be correct PSMs. Despite a complex mathematical vehicle, necessary for statistical guarantees, the original BH procedure is the first and simplest implementation of this approach. However, this procedure relies on a strong assumption: that the scores distributed are p-values, as the null distribution of such values only is known to be uniform[29], at least in theory[30,31]. As such, the BH procedure is the natural tool to control for the FDR when analyzing differential expression, where statistical tests are applied to all putative biomarkers. However, it can also be applied for peptide identification, provided PSM scores can be converted into p-values[32,33].

If no p-value can be determined from the PSM scores, the approach remains valid, but an additional preliminary step is necessary. The purpose of this step is to estimate how PSM scores distribute under the null hypothesis (to keep the subsequent subtraction from the observed distribution feasible). This extension of the BH framework is naturally referred to as "empirical null estimation" (or "Empirical Bayes estimation" when the alternative hypothesis is also accounted for). Related approaches have been used in proteomics for two decades[13,34], and are still under investigation[35]. TDwoC is their quintessence, as it provides a universal, conceptually simple, and easy-to-implement means to derive the distribution of random matches.

To summarize, when decoy sequences are used for empirical null modelling, they must be considered as a whole, essentially as a means to describe the data under the null hypothesis. As this distribution will subsequently be subtracted from the target distribution, it must remain unaltered. Notably, this implies that all the decoys must be accounted for, which seems incompatible with selecting only a subset of them based on how they compete against target sequences.

### 2.2. "Challenging targets" or the competition-based approach

The second approach makes no attempt to elicit the null distribution. It only assumes the existence of a procedure to mimic the features of interest (be they amino acid sequences or quantitative vectors describing hypothetical differential abundances). Each feature thus fictionalized is used to challenge an original feature through pairwise competition. Then, the FDR can be estimated thanks to the overall analysis of all the competition results. Despite regular use in proteomics over 15 years as part of TDC, this approach has only recently been theoretically justified[21,24]. Therefore, it is now tempting to consider it as the post-hoc support for TDC applications that has been sought for more than a decade. Unfortunately, detailed analysis of this so-called knockoff framework, that we will summarize below, appears to partially contradict this view.

Knockoff-based FDR is notably applied in biomarker discovery. It considers as input an array-like quantitative dataset, where each covariate (e.g., a vector of abundances, polymorphism presence/absence across samples, etc.) is a potential biomarker. The first step of the method aims to fabricate a "knockoff" array of the exact same size as the original one. This knockoff array can be pictured as similar to the permuted array used in permutation-based FDR[36], except for two differences: first, the knockoff values are randomly generated from scratch rather than by shuffling the samples. Second, whereas permutation-based FDR relies on an overall description of the null distribution (like BH), knockoff covariates are paired with the originals in order to fit the so-called "exchangeability property"[24]. Respecting this property is challenging, notably when the number of features considerably exceeds the number of samples[25]. Multiple methods have now been described to generate the knockoffs. Figure 1 presents the original procedure, which, despite its complexity, only approximately obeys the exchangeability property. More specifically, the covariance matrix for the joint original and knockoff covariates must read as:

$$\text{cov}(X, \tilde{X}) = \begin{pmatrix} \Sigma & \Sigma - \text{diag}(s) \\ \Sigma - \text{diag}(s) & \Sigma \end{pmatrix} \quad (1)$$

where $X$ and $\tilde{X}$ refers to original and knockoff covariates, respectively; $\Sigma$ is the estimated covariance matrix of $X$; and $\text{diag}(s)$ is a diagonal matrix built upon a vector $s > 0$. To enable sampling of the knockoff covariates, $s$ must be chosen such that $\text{cov}(X, \tilde{X})$ is invertible. In addition, the power of the FDR control procedure depends directly on the coefficients of $s$ being large enough. As a result of this trade-off, tuning $s$ in a high-dimensional setting is challenging. In our view, this clearly illustrates the difficulty of fabricating sufficiently realistic fictional variables. From a more practical viewpoint, TDC-based FDR presents a similar pitfall. As formally defined by He et al.[21], the accuracy of TDC depends on whether the decoy database complies with the "Equal Chance Assumption" during the subsequent competition. Although it is necessary to assume equally probable target mismatches and decoy matches to control the FDR at peptide-level, little is known about the true validity of the assumption. Notably, it has already been reported that instrument mass tolerance filters[32] or gene expression filters[33] can affect the correctness of this assumption. Therefore, many other experimental details may have similar impacts. Likewise, it has recently been reported that the diversity of decoy fabrication methods (see next section) or competition modes[37,38] affects the liberal/conservative behavior of the FDR estimate, and we postulate that the knockoff framework may be useful in explaining these effects.

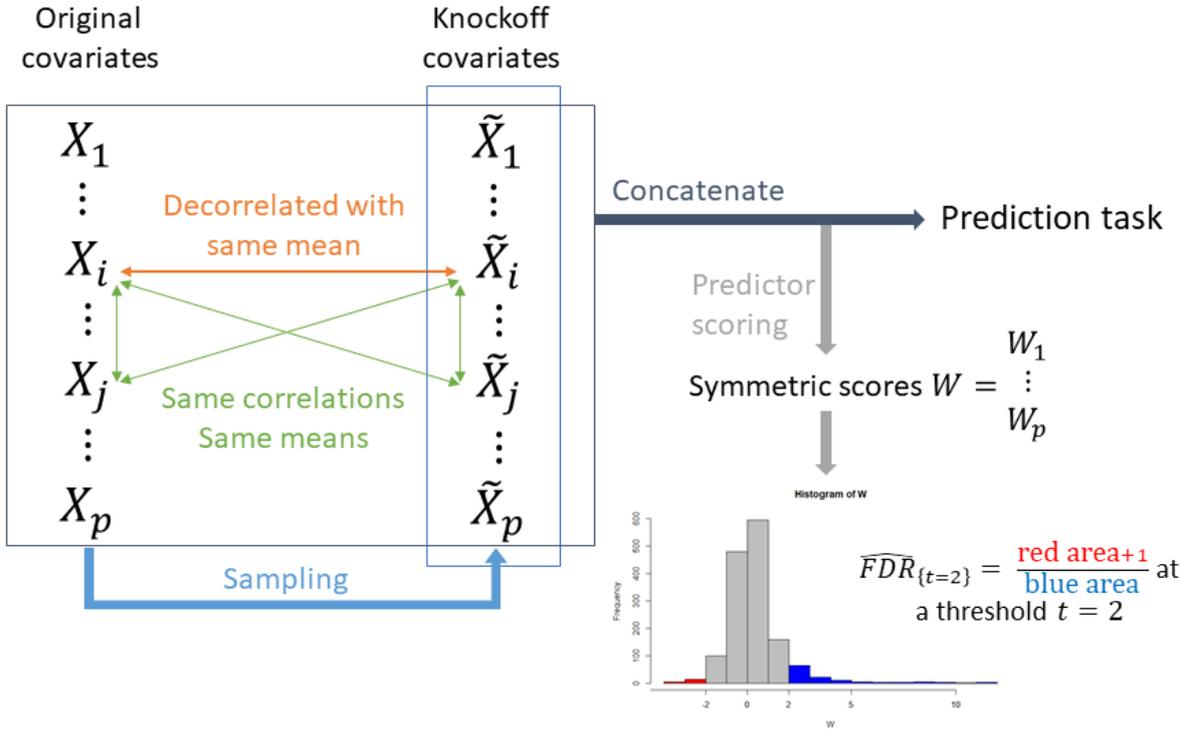

Figure 1: General framework of FDR control with knockoff variables (p denotes the number of covariates).

After generating the knockoffs, a competition is organized between each pair of original and knockoff covariates. This competition produces a symmetric score $W_i$ measuring which of the $i$-th original/knockoff covariates is the best suited to a dedicated task (for differential analysis, this task is classically predicting biological status). The more negative the score, the more it indicates the knockoff outperformed the original feature; and conversely, the more positive the score, the more it indicates the original feature outperformed the knockoff (zero corresponds to a draw between the two). Finally, FDR control at level α is achieved as follows: the minimum score threshold $t$ is chosen such that the ratio $(K_t + 1)/O_t$ is lower than $\alpha$, where $K_t$ (resp. $O_t$) is the number of original/knockoff pairs with relative scores lower than $-t$ (resp. larger than $t$), see Figure 1. All original covariates with a relative score above $t$ are selected. At this point, the parallel with TDC is easy to draw. First, as argued by Keich et al.[39], searching a single concatenated database (the original TDC formulation) or two separate databases followed by a competition step is equivalent. Thus, TDC implements a pairwise competition between the target and decoy sequences best matching each spectrum. Second, the FDR control formula is the same as that recently adopted for peptide identification[21,39,40] (as opposed to the original formula[15,16]). However, at this stage, a small difference emerges. According to the knockoff procedure, the TDC scores (i.e., the $W_i$s) should be obtained using an anti-symmetric function (meaning that $f(x,y) = -f(y,x)$). In practice, the following formula is used:

$$\text{Score}(PSM_i) = W_i = f(Z_i^t, Z_i^d) = \text{sign}(Z_i^t - Z_i^d) \times \max(Z_i^t, Z_i^d) \qquad (2)$$

where $PSM_i$ is the PSM associated with the $i$-th spectrum, and $Z_i^t$ and $Z_i^d$ refer, respectively, to the best target and decoy scores w.r.t. this spectrum. Using this formula, the score complies with the knockoff framework. However, whereas one should retain a null score when $Z_i^t = Z_i^d$, in classical TDC implementations, $Z_i^t$ is often returned instead, thereby breaking the anti-symmetry property. Although justified by practical considerations (if a random amino acid sequence appears to be equivalent to an existing peptide, it should not prevent the peptide from being identified), it may hamper the overall statistical correctness of the procedure in practice.

To summarize, although the parallels between TDC and knockoffs are strong, three discrepancies should be kept in mind: the statistical framework cannot handle FDR control at PSM level; the TDC score may not be perfectly symmetrical; and the procedure used to shuffle amino acids does not guarantee compliance with the exchangeability property.

To conclude on the competition-based approach, introducing a pairwise competition step between the fictional and the original variables corresponds to a significant change in the FDR control procedure: it is supported by a distinct mathematical theory and requires specific implementations to work. This calls for caution in the proteomics community, where switching between frameworks has essentially been taken to mean the target and decoy databases are concatenated (competition-based approach) or kept separate (null-based approach).

## 3. Perspectives inspired by this history

As we have seen, FDR control is possible using two distinct and orthogonal mathematical theories. When applied to peptide identification, both can rely advantageously on decoy generation. However, the role of the decoy database fundamentally differs, depending on whether the method applied relies on BH and its empirical null extensions (then, decoys are used to "describe the mismatches") or whether it relies on a competition-based approach (then, decoys must adequately "challenge the target sequences"). In our view, this fundamental distinction is worth highlighting, as it provides interesting cues to analyze and improve the tools used for peptide identification.

First, TDC improvements can be expected if we better acknowledge the requirements for generation of knockoffs. Notably, the discrepancy between the TDC scoring function and knockoff requirements (see above) should encourage the investigation of alternative scoring systems, for instance leveraging the difference between the best decoy and best target scores. Following this trend, Emery et al.[41,42] proposed the use of multiple decoy database searches and an anti-symmetric score based on the proportion of best decoy matches that were outperformed by the best target matches, similar to other theoretical variations on the knockoff framework[23,26]. Knockoffs are also inspiring for decoy generation. Essentially, a good knockoff amounts to a random variable following the null hypothesis given its original counterpart. However, this distribution is more difficult to define for mismatching amino acid sequences than for differential expression (where knockoffs easily apply). In fact, this question pervades the problem of decoy fabrication, as a number of methods (reverse, shuffle, De Bruijn, etc.)[43,44] have been proposed to allow a trade-off between unrealistic and target-like decoys. However, no consensus has yet emerged. In the knockoff framework, this question boils down to balancing the fit to the exchangeability property, and the assumptions about the null hypothesis. A number of original and inspiring tools have been proposed to achieve this balance, such as the Deep Knockoff approach[45]. This suggests using a generic distribution learner, like a variational autoencoder, with a loss function penalized by the similarity between the target and decoy scores over the training spectra.

Second, and more pragmatically, swapping the default FDR control methods and the use-cases highlights the pros and cons of each approach. We applied BH to peptide identification data[32,33] and conversely, for differential analysis, we replaced BH by knockoffs[46]. Our results concurred and showcased the considerable instability of the competition-based approach relative to null-based approach. This instability is presumably linked to random fluctuations during decoy generation, which directly influence the results of pairwise competitions. In contrast, the overall description provided by the null distribution should be less sensitive to random variations. To cope with these fluctuations, it has been proposed to run multiple TDCs and to average the target and decoy counts to estimate the FDR[39] (an approach that should not be confused with the multiple knockoff approaches mentioned above[26]). However, any such averaging strategy comes at an extra computational cost, which is not required when applying the null-based approaches.

This difference in stability also provides a possible explanation for the results presented by Madej et al.[1] More specifically, the authors proposed to apply the Common Decoy Distribution (CDD) in two distinct ways: BH-CDD (Benjamini-Hochberg CDD) and PP-CDD (PeptideProphet CDD). Technically speaking, both implementations amount to a null-based approach, as both use the CDD as an empirical null distribution. The BH-CDD copes with the lack of available p-values by relying on a simple empirical null model to estimate an overall FDR, whereas PP-CDD – following PeptideProphet approach[13,34] – uses a joint modelling of the null and alternative hypotheses to estimate the local FDR distribution[12]. Madej et al.[1] reported that, compared to BH-CDD, PP-CDD produces FDR estimations with greater variance. Considering the lack of stability intrinsic to the random generation of fictional variables (whether knockoffs or shuffled sequences), we believe that the minimal data-dependency of BH-CDD (in contrast to PP-CDD, which requires an additional distribution to be estimated from the data) explains its relative stability. As for the reported over-conservativeness of BH-CDD, refinements using the PIT (Percentage of incorrect target PSMs[17], a.k.a. $\pi_0$ )[19,30] are known to be efficient.

Regardless of the strategy applied (PP-CDD or BH-CDD), the distinction between the null- and competition-based approaches sheds an interesting light on the proposal presented by Madej et al.[1]. Indeed, it constitutes a middle-of-the-road approach, as the construction of the CDD involves pairwise competitions, whereas the FDR estimate relies on the empirical null paradigm. Is such an in-between theoretically supported? So far, it does not appear to be, to the best of our knowledge. However, the experimental results reported seem at least partly compliant with a well-calibrated FDR control procedure. This can be explained in a number of ways.

First, when conducting a peptide identification task with most available database search engines, the null hypothesis is not that of random mismatches, but that of the best mismatches over the entire database (most search engines only return the few best-scoring PSMs, whether targets or decoys, and classically, only the best one is retained). In other words, even a decoy-only search entails a kind of competition step (in the sense that it amounts to taking the highest of several scores), regardless of its subsequent use (with or without competition against target results). Potentially, after the competition involving the entire decoy database, adding another level of competition with the best target does not significantly alter the distribution. Of course, it should yield a more conservative CDD as only some decoys (those defeating the targets) are considered to describe the null distribution. However, let us recall that depending on how the databases are filtered[32,33], the TDC-based FDR can be anti-conservative. The two errors could thus compensate for each other.

An alternative explanation can also be presented: although we currently lack theoretical results, a mixed approach may still hold. It should be remembered that the concept of TDC emerged more than 10 years earlier than the theoretical framework of knockoffs. Similarly, future statistics studies may give grounds for mixing strategies rooted in computational proteomics, where the overall null distribution is determined from a series of pairwise competitions. Although taking a slightly different path, some theoretical attempts to bridge the gap between knockoff- and p-value-based FDRs are already emerging[46,47].

In any case, the results reported by Madej et al.[1] are in line with those summarized above: To date in a proteomics application context, competition-based approaches do not appear to us as mature as their null-based counterparts. Owing to the random fluctuations inherent to the generation of fictional features, and the difficulty of ensuring that this generation remains compliant with the mathematical constraints of the underlying theory (which, broadly speaking, is essential to comply with the Equal Chance Assumption), null-based FDR controls should be preferred. More precisely, we tend to promote the BH control, as it is the most stable and the least computationally demanding. Unfortunately, it cannot easily be used with many search engines that do not directly provide p-values. In this context, the most striking application of CDD we envision is a universal method to convert search engine scores to p-values. By definition, p-values distribute like their theoretical quantiles, therefore the empirical quantiles of the scores attributed by a search engine to a CDD provide a correspondence table between the scores and the p-values for the search engine in question. Subsequent application of the BH procedure then becomes straightforward. In addition to the gains in terms of accuracy and simplicity for FDR control, this would give a common ground to simplify the comparison of the various search engines available in the literature.

# 4. Conclusions

In conclusion, the objective underlying the definition of a Common Decoy Distribution (CDD) is as interesting as the new applications it makes possible. However, its practical use raises many questions, mainly because it requires a distinction to be made between two uses of decoy sequences when controlling for the FDR: either to challenge the target sequences in a pairwise competition setting, or to refine the description of the null hypothesis (a.k.a. target mismatches). Doing so in a proteomics context is difficult, since both trends have been tightly intertwined over the past twenty years. The reason for this intermingling is that our community adopted the FDR concept in a progressive and sinuous manner, fueled by a mix of empirical considerations and concomitant theoretical results. Fortunately, the recent advent of knockoff theories is insightful in this respect. In this context, the proposal from Madej et al.[1] constitutes a milestone encouraging further investigations in multiple directions. First, casting the CDD principle into a fully empirical null framework (by removing the competition step during CDD construction) would produce a tool that could be extensively used to convert search engine scores into p-values. Second, by providing the means to explore a range of target-decoy strategies, CDD could well become a practical tool to refine our current approaches to FDR control (e.g., stability studies, anti-symmetric scores, averaging results from multiple small databases vs. using a single large one, $\pi_0$ estimate, etc.). Finally, in the longer term, such explorations should contribute to more theoretical investigations; notably attempts to bridge the gap between the statistical rationales of the null- and competition-based approaches, possibly leading to the emergence of a unified theory.

# 5. Declarations

**Author contributions**
LE contributed to the literature survey, handled the statistical aspects, and designed the figure. TB provided the original idea, contributed to the literature survey and drafted the manuscript. Both authors finalized the manuscript and approved the submitted version. The authors wish to thank Maighread Gallagher (TWS Editing) for the proof-reading service.

**Funding**
This work was supported by grants from the French National Research Agency: ProFI project (ANR-10-INBS-08), GRAL project (ANR-10-LABX49-01) and MIAI @ Grenoble Alpes (ANR-19-P3IA-0003).

**Conflict of interest**
The authors declare no competing financial interest.